\documentclass[conference, letterpaper]{IEEEtran}

\newcommand{\Gateway}[1]{GW#1}
\newcommand{\LearningRate}[1]{\ensuremath{\alpha}#1}
\newcommand{\DiscountFactor}[1]{\ensuremath{\gamma}#1}
\newcommand{\AcceptedThreshold}[1]{\ensuremath{\tau}#1}
\newcommand{\Observation}[1]{\ensuremath{o_t}#1}

\newcommand{\reward}[1]{\emph{reward#1}}
\newcommand{\action}[1]{\emph{action#1}}

\newcommand{\state}[1]{\emph{state#1}}

\newcommand{\observation}[1]{\emph{observation#1}}

\usepackage[usenames, dvipsnames]{color}

\usepackage{amsmath}
\usepackage{mathtools}
\usepackage{amssymb}

\usepackage{dsfont}
\usepackage{paralist}
\usepackage{footnote}
\usepackage{array}
\usepackage{ragged2e}
\usepackage{bm}
\usepackage{textcomp}
\usepackage[table]{xcolor}

\usepackage{subcaption}
\usepackage{mathtools}
\usepackage{algorithmic}

\usepackage[linesnumbered]{algorithm2e}

\ifCLASSINFOpdf
  \usepackage[pdftex]{graphicx}
  \graphicspath{{../figures/}}
\else
\fi

\begin{document}


%
\title{Adapting Sampling Interval of Sensor Networks Using On-Line Reinforcement Learning}

\author{\IEEEauthorblockN{Gabriel Martins Dias, Maddalena Nurchis and Boris Bellalta}
\IEEEauthorblockA{~\\Department of Information and Communication Technologies\\ 
Universitat Pompeu Fabra, Barcelona, Spain\\
Email: \{gabriel.martins, maddalena.nurchis, boris.bellalta\}@upf.edu}
}

\maketitle

\begin{abstract}

Monitoring Wireless Sensor Networks (WSNs) are composed of sensor nodes that 
report temperature, relative humidity, and other environmental parameters. 
The time between two successive measurements is a critical parameter to set 
during the WSN configuration because it can impact the WSN's lifetime, the 
wireless medium contention and the quality of the reported data. 
As trends in monitored parameters can significantly vary between scenarios and 
within time, identifying a sampling interval suitable for several cases is also 
challenging. 
In this work, we propose a dynamic sampling rate adaptation scheme based on 
reinforcement learning, able to tune sensors' sampling interval on-the-fly, 
according to environmental conditions and application requirements. 
The primary goal is to set the sampling interval to the best value possible so 
as to avoid oversampling and save energy, while not missing environmental 
changes that can be relevant for the application. 
In simulations, our mechanism could reduce up to $73\%$ the total number of 
transmissions compared to a fixed strategy and, simultaneously, keep the 
average quality of information provided by the WSN. 
The inherent flexibility of the reinforcement learning algorithm facilitates 
its use in several scenarios, so as to exploit the broad scope of the Internet 
of Things.

\end{abstract}


\section{Introduction}

The broad adoption of wireless sensor nodes for environmental monitoring 
purposes happened not only because of their capacity of sensing and 
transmitting (via radio) several environmental parameters but mainly thanks to 
their low production costs. 
In fact, their low production costs are a result of limited resources, such as 
battery and memory, which have boosted lots of research work aimed at extending 
their lifetime without affecting their most valuable asset: the sensed data.

In monitoring Wireless Sensor Networks (WSNs), the sensor nodes' sampling 
interval is crucial for generating high-quality data. 
The data quality is reduced when the interval between two measurements is not 
sufficiently short to report significant changes in the monitored parameters, 
or if it is too brief and report several similar (and, consequently, 
unimportant) successive values. Furthermore, a sampling interval set under some 
conditions may occasionally become too short (or too long) within time, due to 
the environment's evolution. Additionally, besides the data quality, the 
sampling interval affects the wireless medium access and stirs up network 
congestion, end-to-end delays, and sensor nodes' energy consumption. 
In conclusion, the sampling interval configuration must focus on the trade-off 
between data quality and resource savings. 

In this work, we propose an on-line sampling interval adaptation scheme. 
Our mechanism relies on real-time analysis of the data produced by sensor nodes 
to dynamically adapt the WSN operation to the current environmental conditions. 
Our primary goal is to guarantee the minimum number of transmissions needed to 
avoid losing valuable environmental data. 
To achieve that, we aim to keep the maximum difference between two consecutive 
measurement values below a threshold, which is set according to the application 
needs.

We formally represent the scenario of a monitoring WSN through the 
reinforcement learning model and apply a Q-Learning algorithm to learn the most 
suitable sampling intervals under different conditions, without an a-priori 
model of the environment's evolution. 
The learning agent may be placed in the cloud, a central gateway or each 
sensor, according to the resource constraints. 
The scope of our work is to evaluate the algorithm and observe factors that 
impact its performance, whereas analyzing the performance difference between  
centralized and distributed solutions is left for future work.

The rest of the paper is organized as follows:
Section~\ref{sec:reinforcement-learning} describes the basics of 
reinforcement learning;
Section~\ref{sec:related-work} enumerates related works that adopted 
reinforcement learning to optimize sensor networks at different layers, and 
those that focused on avoiding unnecessary transmissions without losing 
important changes in the environment;
Section~\ref{sec:adaptive-sampling} formulates the problem of adapting 
the sensor nodes' sampling interval as a reinforcement learning problem;
Section~\ref{sec:synthetic-scenarios} explains how we generated controlled 
scenarios and experimental results;
Section~\ref{sec:real-scenarios} shows experimental results of the 
reinforcement learning technique applied to real datasets;
Section~\ref{sec:conclusion} shows our conclusions and ideas for future work.

\section{Background - Reinforcement Learning}
\label{sec:reinforcement-learning}

Reinforcement Learning (RL) is a machine learning technique
that allows an \emph{agent} to determine automatically a system's optimal 
behavior to achieve its goal~\cite{Sutton:1998:IRL:551283}.
Such an optimal behavior is based on the positive and negative feedbacks 
received from the environment after taking certain \action{s}. 
Assuming that interactions between the agent and the environment occur at a 
sequence of discrete time instant $t$, an RL model is defined by:
\begin{itemize}
 \item a set of possible \observation{s} $O$ that the agent may make, such that 
$o_t \in O$ is the observation made at time $t$;
 \item a set of \state{s} $S$, such that the \state{} $s_t \in S$ is observed 
at time $t$;
 \item a set of \action{s} $A$, such that the \action{} $a_t \in A$ is 
taken at time $t$;
 \item a state transition function $T(s_t, a_t, s_{t+1})$ that calculates the 
probability of making a transition from $s_t$ to $s_{t+1}$ after performing 
$a_t$; and
 \item a set of rules that determine the scalar immediate \reward{} $r_{t+1} = 
R(s_t, a_t)$, which scales the goodness of taking $a_t$ in $s_t$.
\end{itemize}

Each \state{} should satisfy the Markov property\footnote{RL can also be 
applied to cases that do not satisfy the Markov 
property~\cite{Sutton:1998:IRL:551283}}, that is, 
to be independent of any \state{} or \action{} previous to time $t$.
An RL agent aims to obtain the maximum long-term \reward{} for a 
Markov Decision Process environment, even when the model of the 
environment is unknown or difficult to learn. 
The strategy adopted to maximize the long-term \reward{} defines the 
agent's way of behaving at a certain time and is called a \emph{policy}.

\subsection{Q-Learning}

Q-Learning is an RL algorithm that does not depend on a state transition 
function to work. 
More precisely, the algorithm relies on an optimal action-value function 
$Q(s, a)$, which value is the estimated \reward{} of executing 
$a$ in $s$, assuming that the agent will always follow the 
\emph{policy} that provides the maximum long-term \reward{}.

At any state $s_t$, a selected action $a_t$ determines the transition to the 
state $s_{t+1}$ and the value associated to the pair ($s_t$, $a_t$) is updated:

\begin{equation}
\begin{split}
Q_{t+1}(s_{t},a_{t}) = &~\LearningRate{}\left(r_{t+1} + 
\DiscountFactor{}\max_{a}Q_t(s_{t+1},a)-Q_t(s_t,a_t)\right)
\\ &+ Q_{t}(s_t, a_t)
\text{,}
\end{split}
\end{equation}
where $s_{t+1}$ and $r_{t+1}$ are the \state{} and \reward{}, respectively,
obtained after performing $a_{t}$ in $s_{t}$, the \emph{learning rate} 
$\LearningRate{} \in [0, 1]$ is a positive step-size parameter, and the 
\emph{discount factor} $\DiscountFactor{} 
\in [0, 1]$ is used to determine the weight of future \reward{s}.
If $\DiscountFactor{} = 0$, the agent will behave so as to maximize its 
immediate \reward{}, even if this would imply a lower long-term return.

By visiting several times each ($s$, $a$) pair, the agent learns which is the 
action that gives the best long-term \reward{} in each state.
Hence, if the number of \state{s} is high, the algorithm takes longer
and requires more data to find the best \action{} for each \state{}, i.e., to 
\emph{converge}. 
Therefore, it is critical to have a concise representation of the 
environment, thus to define the set of \state{s} according to the goals of the 
algorithm and do not include unnecessary information.
In short, the set of \state{s} should illustrate only and all the
characteristics that are relevant for the problem under consideration.

\section{Related Work}
\label{sec:related-work}

Several works have already adopted reinforcement learning techniques at various 
layers to improve wireless networks' performance.
In~\cite{nurchis2011self}, the authors proposed a self-adaptive routing 
framework for wireless mesh networks.
Using the Q-Learning algorithm, it was possible, in runtime, to select the 
most proper routing protocol from a pre-defined set of options and successfully 
increase the average data throughput in comparison to static techniques.
Other examples of the use of reinforcement learning techniques in sensor 
networks include locating mobile sensor nodes~\cite{li2012dynamic} and 
aggregating sensed data~\cite{Mihaylov2010}. 

WSNs are mainly composed of wireless sensor nodes that make measurements and 
transmit them to a Gateway (\Gateway{}).
There are many solutions to reduce their number of transmissions, which 
include clustering, data aggregation, and data prediction. 
In~\cite{LeBorgne2007}, the authors suggest that future measurements can be 
predicted by sensor nodes and \Gateway{s}.
Therefore, sensor nodes only transmit a measurement if they observe that the 
prediction is not correct, i.e., the real measurement differs by more than a 
certain threshold from the predicted value.
The success of this technique highly depends on the capacity of the sensor 
nodes to compute efficient prediction methods that will accurately predict
future values.
However, sensor nodes usually have very limited computing capacities and must 
rely on \Gateway{s} to regularly generate and transmit new predictions.
Moreover, \Gateway{s} and sensor nodes must share the same knowledge, which 
requires additional control messages and reduces the benefit of decreasing 
measurement transmissions, especially if predictions are not sufficiently 
accurate.

In~\cite{Malik2011}, the authors propose an approach to answering queries in 
\Gateway{s} without fetching the data directly from sensor nodes.
The Principal Component Analysis method was used to analyze historical data 
and select only the sensor nodes that measured most of the variance observed in 
the environment.
This technique reduced the workload of the sensor nodes and reduced up 
to $50\%$ the number of transmissions, according to the results obtained from 
experiments in real testbeds. 
However, the authors do not define how the environment's evolution would be 
addressed.
For example, if the temperature varies more often during the day,
it would be necessary more measurements from more sensor nodes during these 
hours to build datasets that reliably describe the environment.

Our work does not necessarily
rely on the computational capacity of sensor nodes or \Gateway{s} 
because the incorporation of WSNs into the IoT allows the use of external 
entities and cloud computing services that can perform powerful machine 
learning 
techniques over sensed data if needed~\cite{Dias2016c}.
To the best of our knowledge, this is the first approach that dynamically adapts 
the 
sampling interval of the sensor nodes based on a reinforcement learning 
technique.


\section{Adaptive sampling as a reinforcement learning problem}
\label{sec:adaptive-sampling}

In this Section, we formulate the adaptive sampling interval problem as a 
reinforcement learning problem.
For this work, we take as a reference the real scenario of a WSN with several 
nodes measuring temperature values in an office~\cite{IntelLabData:2004:Misc}. 
From this real dataset, we use a sub-set of measurements in the preliminary 
analysis presented in Section~\ref{sec:states}, and a different sub-set in the 
performance evaluation in Section~\ref{sec:real-scenarios}
(missing values were interpolated and added to a small white noise). 
Moreover, as the sensors of this scenario were set to sample temperature 
nearly every $30$ seconds, we set this as the shortest sampling interval and 
let the range of possible sampling intervals be\begin{inparaenum}[(i)]
 \item \textbf{$\mathbf{30}$ seconds};
 \item \textbf{$\mathbf{60}$ seconds};
 \item \textbf{$\mathbf{120}$ seconds}; or
 \item \textbf{$\mathbf{240}$ seconds}.
\end{inparaenum}

A valuable adaptive sampling algorithm should systematically set up the most 
proper sampling interval so as to guarantee the best quality-resource trade-off 
under the current environmental conditions. 
As for the quality, we define the goal of the agent in terms of an 
\textit{accepted threshold} \AcceptedThreshold{}.
The algorithm should avoid that the absolute difference between consecutive 
measurements exceeds a pre-defined value (\AcceptedThreshold{}), which 
is configured according to the monitoring application requirements. 
Meanwhile, higher sampling intervals are preferred to reduce the number of 
transmissions and, consequently, the energy consumption in sensor nodes.

In our experiments, we consider the number of transmissions as a general 
measure of resource optimization, which is also valuable in scenarios with 
energy constraints.
For space limitation, we can not include other metrics that may be more 
relevant 
in specific scenarios, such as the energy saved by reducing report 
transmissions 
or by increasing the idle time of the sensor nodes. 

\subsection{Observations}

First, we define wireless sensor nodes as the source of the observations made 
by 
an agent.
Observations may vary, among other parameters, between temperature, 
relative humidity and solar radiation.
In our scenario example, an \emph{observation} \Observation{} is the 
temperature measured at time $t$.

\begin{table}[t]
\centering
\def\arraystretch{1.0}
\rowcolors{2}{white}{gray!25}
\begin{tabular}{
	>{\arraybackslash}m{2.8cm}
	>{\arraybackslash}m{5.15cm}
	}
  \cellcolor{gray!50}\textbf{Factor} & 
  \cellcolor{gray!50}\textbf{Description} \\
  \textbf{Quality} ($q$) & $q \triangleq |o_t - o_{t-1}| \leq 
\AcceptedThreshold{} \therefore 
q \in \{\text{true}, \text{false}\}$ \\ 
  \textbf{Hour of the day} ($h$) & $h \in [0, 1,\ldots, 23]$ \\ 
  \textbf{Is it working hour?} ($\omega$) &$\omega \triangleq h \geq 
7~\text{and}~ h \leq 18 \therefore \omega \in \{\text{true}, \text{false}\}$ \\
  \textbf{Day of the week} ($d$)& $d \in \{\text{Monday},~
\text{Tuesday},~\ldots,~\text{Sunday}\}$ \\ 
  \textbf{Is it weekend?} ($e$) & $e \triangleq d \in 
\{\text{Sat.},~\text{Sun.}\} \therefore e \in \{\text{true}, 
\text{false}\}$ \\ 
  \textbf{Sampling interval} ($s$) & $s \in \{30, 60, 120, 240\}$ seconds \\
  \textbf{Node ID} ($i$) & Individual sensor node identification. \\ 
   \hline
\end{tabular}
\caption{Factors that can impact the quality of measurements.}
\label{tab:factors}
\end{table}

\subsection{States}
\label{sec:states}

To properly define the set of possible \state{s}, we make a preliminary 
evaluation of part of the data collected by the WSN, to identify 
characteristics that have a high correlation with our goal.
Such characteristics are transformed into predictors, and a Random 
Forest~\cite{Breiman2001} is built to classify in which periods of time 
each sampling interval would make consecutive measurements that differ by less 
than \AcceptedThreshold{}.

In this process, we use the data from three different sensor nodes sampled 
every $30$ seconds for three days.
To simulate different sampling intervals, we removed intermediate measurements 
and, after analyzing the generated data, the value of \AcceptedThreshold{} 
was set to $0.02^{o}$C.
Using this value, a sampling interval of $120$ seconds would be sufficient to 
observe a difference of less than \AcceptedThreshold{} in nearly one-half of 
the time.
Furthermore, sampling intervals of $30$, $60$ and $240$ seconds would be 
sufficient to observe a difference of less than \AcceptedThreshold{} in, 
respectively, approximately $73.2\%$, $59.2\%$ and $26.1\%$ of the time.
Note that we do not expect to measure fast enough to make all measurements 
differ by less than \AcceptedThreshold{}, but we know that there are many 
cases in which sensors do not have to sample every $30$ seconds to
achieve it.

Finally, we annotate each measurement with the characteristics shown in 
Table~\ref{tab:factors} and build a Random Forest to observe what are the most 
relevant factors to predict if a measurement will differ by less than 
\AcceptedThreshold{} from the previous one (i.e., $q = \text{true}$).
Based on the results obtained using the Random Forest algorithm and considering 
the importance of keeping a small number of \state{s}, we defined the set of 
\state{s} for our RL model as a combination of quality, sampling interval and a 
verification of whether it is a working hour or not (see 
Table~\ref{tab:factors}) .
Thus, each \state{} is defined by a tuple: $\{q, s, \omega\}$.
In short, as we are looking for \state{s} that will be used by different 
sensor nodes, we ignored any factor that was less important than the node ID to 
predict the quality of the measurements and those that would represent a 
significant increase in the number of states, such as the hour of the day.

\subsection{Actions and transitions}

In the adaptive sampling interval problem, \action{s} are used to control the 
sensor nodes' sampling interval.
An \action{} can be specific, such as ``set the sampling interval to 30 
seconds'', or more abstract, like ``increase the sampling interval'' requiring 
the new sampling interval to be calculated based on the last value set.
To avoid abrupt changes provoked by occasional outliers and noise in 
the data, we adopt only ``smooth'' \action{s}, i.e., they only move to 
neighboring \state{s}.
Therefore, an \action{} $a$ can take one of the following 
values:\begin{inparaenum}[(i)]
 \item \textbf{increase the sampling interval};
 \item \textbf{keep the sampling interval}; or
 \item \textbf{reduce the sampling interval}.
\end{inparaenum}

\subsection{Reward}

A \reward{} is a mathematical representation of the gains obtained after 
reacting to the environment with a particular \action{}.
In our case, after changing the sampling interval to a new value, 
while in $s$.
As the reward defines the target of the algorithm,
in our problem, it should ensure that the difference between consecutive 
measurements is less than \AcceptedThreshold{}, while not oversampling.

The algorithm adopted for the \reward{} is based on the rate of transmissions 
avoided.
For instance, if the sampling interval is $120$ seconds, the sensor node is 
transmitting four times less than if it was $30$ seconds.
In this case, therefore, the original \reward{} would be set to four.
Then, if the absolute difference between two consecutive measurements 
($\delta$) is less than \AcceptedThreshold{}, we assume that the sampling 
interval is small enough to avoid losing significant changes in the environment 
and take the original \reward{}.
If $\delta$ is smaller than one-half of \AcceptedThreshold{}, the sampling 
interval might be doubled, so we multiply the original \reward{} by 
$\mathbf{1.5}$.
Otherwise, if $\delta$ is greater than \AcceptedThreshold{}, the 
sampling interval is too long, and the reported data may be missing important 
changes in the environment.
In this case, we multiply the original \reward{} by$~\mathbf{-1}$.

\section{Synthetic scenarios}
\label{sec:synthetic-scenarios}

To check the feasibility of using RL as a means of intelligently adapting 
sampling intervals, we simulate its use in artificial and realistic scenarios.
In this Section, we present the simplest scenarios, using synthetic data 
with evident characteristics, such as vast and significant (versus small and 
negligible) variations in a short period.
Having control over the data characteristics allows us first to verify if 
the RL algorithm decides for the most proper \action{s} in different scenarios.
Later, we analyze the impact of the values of \emph{learning rate} 
\LearningRate{} and \emph{discount factor}~\DiscountFactor{} on the time 
that the agent needs to reach the most proper \state{} when it occurs.

\begin{savenotes}
\begin{table}[t]
\centering
\def\arraystretch{1.0}
\rowcolors{2}{white}{gray!25}
\begin{tabular}{
	>{\centering\arraybackslash}m{1cm}
	>{\centering\arraybackslash}m{1cm}
	>{\centering\arraybackslash}m{1.4cm}
	>{\centering\arraybackslash}m{1.3cm}
	>{\centering\arraybackslash}m{1.9cm}
	}
  \cellcolor{gray!50}\textbf{Learning rate \LearningRate{}} & 
  \cellcolor{gray!50}\textbf{Discount factor \DiscountFactor{}} & 
  \cellcolor{gray!50}\textbf{Convergence time (s)} &
  \cellcolor{gray!50}\textbf{\% of wrong decisions} &
  \cellcolor{gray!50}\textbf{\% of measurements over 
\AcceptedThreshold{}\footnote{\label{footnote:tau}Without considering datasets 
\emph{Controlled 30} and \emph{Controlled 240}.}}\\ 
0.9 & 0.1 & 1013.00 & 4.79 & 2.31 \\ 
  0.8 & 0.2 & 1050.32 & 6.11 & 4.38 \\ 
  0.8 & 0.1 & 1088.01 & 8.36 & 1.80 \\ 
  0.8 & 0.4 & 1163.05 & 13.46 & 8.80 \\ 
  0.8 & 0.5 & 1640.30 & 12.82 & 10.79 \\ 
  0.9 & 0.2 & 1920.57 & 11.57 & 3.65 \\ 
  0.5 & 0.2 & 18330.53 & 25.39 & 10.89 \\ 
  0.9 & 0.4 & 18457.82 & 19.90 & 4.63 \\ 
  0.6 & 0.1 & 21922.80 & 18.47 & 5.47 \\ 
  0.9 & 0.5 & 23512.98 & 20.68 & 4.34 \\ 
   \hline
\end{tabular}
\caption{Simulations over the \emph{Controlled} datasets: average convergence 
times and percentage of wrongly taken decisions by each combination of 
\LearningRate{} and \DiscountFactor{}. These values are sorted by the lowest 
convergence times.}
\label{tab:mean-controlled}
\end{table}
\end{savenotes}  

\subsection{Fixed expectations}

We generated six synthetic scenarios in which we have control about the 
sampling intervals that the algorithm should set.
In these datasets, the difference between consecutive measurements is 
proportional to \AcceptedThreshold{}.
For instance, if the difference between two consecutive measurements made over 
a period of $60$ seconds is always smaller than \AcceptedThreshold{}, setting 
the sampling interval to $30$ or $60$ seconds is sufficient to satisfy the 
requirements of quality.
However, setting the sampling interval to $60$ seconds would be preferred, 
because it reduces the number of transmissions, in comparison with the $30$ 
seconds interval.

In the \textbf{Controlled 30} dataset, the difference between subsequent 
measurements made over an interval of $30$ seconds between each other is always 
$110\%$ of \AcceptedThreshold{}.
In practice, even the smallest sampling interval ($30$ seconds) is 
not sufficient to provide measurements in which consecutive values differ 
by less than~\AcceptedThreshold{}.
Therefore, the agent must define the ideal sampling interval as $30$ seconds to 
reduce as much as possible the quality loss. 
Note that in this particular scenario, a difference between subsequent 
measurements higher than the maximum threshold is unavoidable.
Therefore, we do not consider this dataset when reporting the percentage
of measurements over \AcceptedThreshold{}.

In the \textbf{Controlled~60}, \textbf{Controlled~120} and 
\textbf{Controlled~240} datasets, the difference between subsequent 
measurements made at an interval of $30$ seconds between each other is 
respectively $47.5\%$  $23.75\%$ and $10\%$ of \AcceptedThreshold{}.
Hence, $60$, $120$ and $240$ seconds are respectively the 
largest possible %
sampling intervals such that the sensor node will never report a difference 
greater than \AcceptedThreshold{}.
Hence, the agent must define the ideal sampling interval respectively to 
$60$, $120$ and $240$ seconds in each scenario. 
Note that, antagonistically to the \emph{Controlled 30} dataset, in 
\emph{Controlled 240}, successive measurements never have an absolute 
difference higher than the maximum threshold.
Therefore, we also do not consider this dataset when reporting the percentage
of measurements over \AcceptedThreshold{}.

To observe the impact of \LearningRate{} and \DiscountFactor{} in the decisions
taken by the agent, we observe how long the Q-Learning algorithm takes to 
decide for the correct sampling interval.
We call this period \emph{convergence time}, and 
we assume that the agent has converged to a final value if the sampling 
interval does not change in, at least, $75\%$ of the future decisions.
The reported values are the average of all considered scenarios.

In our simulations, the average convergence time was less than $2000$ seconds 
(around $33$ minutes) only in six cases and nearly ten times longer in the
remaining. 
We highlight, as WSNs are usually long-term deployments that last for 
months (or years), the period of one day (or less) spent to find the most 
proper sampling intervals represents less than $1\%$ of their average lifetime.
Table~\ref{tab:mean-controlled} shows the combinations with the ten lowest 
average convergence times, the percentage of times that the agent took a 
wrong decision (using the expected sampling interval as a reference) and the 
percentage of consecutive measurements that differed by more than 
\AcceptedThreshold{}.
Half of these combinations had high
\LearningRate{} (i.e., $\LearningRate{} \in \{0.8, 0.9\}$), and low 
\DiscountFactor{} (i.e., $\DiscountFactor{} \in \{0.1, 0.2\}$),
which means that the agent performs better when its decisions
are mostly based on the current status of the environment, and little 
importance is given to future estimated rewards.
In practice, it shows that if a particular action resulted in high \reward{s} 
in the day before, it would not necessarily result in high \reward{s} in the 
future due to the environment's evolution.

\subsection{Moving expectations}

In real world applications, the environment may constantly be changing and 
evolving, requiring that agents never stop to learn, because there might
not exist an answer that stands forever as the most proper one.
To synthesize these situations, we generated three datasets that are 
combinations of the \emph{Controlled} datasets presented before.
Finally, we simulate four days in which the agent should converge to a 
new value each day, updating its previous belief.
In practice, these scenarios will show how good is the algorithm to keep 
learning from the environment's evolution, even after a decision has been 
already taken.

The sequence of expected sampling intervals varies in each dataset.
In \textbf{Evolving~I}, the sampling interval that satisfies 
\AcceptedThreshold{} evolves in the sequence: $30$ seconds in the first day, 
$60$ seconds in the second day, $120$ seconds in the third day and 
$240$ seconds in the last day.
In \textbf{Evolving~II}, the most proper sequence of sampling intervals is 
$240$, $120$, $60$ and $30$ seconds.
In \textbf{Evolving~III}, the most proper sequence of sampling intervals is 
$60$, $120$, $240$ and $30$ seconds.


To evaluate the impact of \LearningRate{} and \DiscountFactor{} on how fast the 
\mbox{Q-Learning} algorithm can adapt to changes in the environment, we observe 
the 
average convergence time among different days.
Again, the convergence time is defined as the initial period that the agent 
takes to decide for the correct sampling interval and does not change in, at 
least, $75\%$ of the future decisions.
The reported values are the daily average of all considered scenarios.


Table~\ref{tab:median-controlled-variations} shows the three parameter
 combinations that took less than $16$ hours to converge  on every simulated 
day and the respective percentage of the agent's decisions that were wrongly 
taken.
Recall in these--more realistic--scenarios the conditions change every $24$ 
hours.
Therefore, every day the agent revisits \state{s} and updates its
knowledge to set the most proper sampling intervals, which increases
the time necessary to converge: 
on average, at least $4.5$ hours more than in the previous 
simulations.
Once again, a high \LearningRate{} (namely, $\LearningRate{} = 0.9$) 
combined with low \DiscountFactor{} (i.e., $\DiscountFactor{} \in \{0.1, 
0.2, 0.5\}$) is the best option to reduce the average time that the agent takes 
to converge to the most proper sampling interval value, considering that the 
environment is continuously evolving.

\begin{savenotes}
\begin{table}[t]
\centering
\def\arraystretch{1.0}
\rowcolors{2}{white}{gray!25}
\begin{tabular}{
	>{\centering\arraybackslash}m{1cm}
	>{\centering\arraybackslash}m{1cm}
	>{\centering\arraybackslash}m{1.4cm}
	>{\centering\arraybackslash}m{1.3cm}
	>{\centering\arraybackslash}m{1.9cm}
	}
  \cellcolor{gray!50}\textbf{Learning rate} & 
  \cellcolor{gray!50}\textbf{Discount factor} & 
  \cellcolor{gray!50}\textbf{Convergence time (s)} &
  \cellcolor{gray!50}\textbf{\% of wrong decisions} &
  \cellcolor{gray!50}\textbf{\% of measurements over 
\AcceptedThreshold{}}\\ 
0.9 & 0.2 & 18417.80 & 22.78 & 7.56 \\ 
  0.9 & 0.1 & 31420.31 & 37.45 & 3.24 \\ 
  0.9 & 0.5 & 31782.82 & 48.30 & 13.64 \\ 
   \hline
\end{tabular}
\caption{Simulations over the \emph{Evolving} datasets: average convergence 
times and percentage of wrongly taken decisions by each combination of 
\LearningRate{} and \DiscountFactor{}.
The values are sorted by the lowest convergence times.}
\label{tab:median-controlled-variations}
\end{table}
\end{savenotes}

\section{Real world Scenarios}
\label{sec:real-scenarios}

In real world scenarios, the environment is constantly changing, 
and there are external (uncontrolled) factors that impact the measurements.
To simulate that, we adopted real measurements collected during five days 
by five wireless sensor nodes and set $0.02^\text{o}C$ as the value of 
\AcceptedThreshold{}, using the strategy explained in Section~\ref{sec:states}.
These measurements were collected in the same experiment we 
considered to setup the \state{s} in Section~\ref{sec:states}, but now we 
use data from different nodes.

To illustrate the results, we assume that during the first $12$ hours, the 
Q-Learning algorithm ``calibrates'' the action-value function, i.e., it 
tries to visit all state-action pairs to estimate the long-term \reward{s}
that each \action{} would provide in each \state{}.
Therefore, we consider only the results observed in the last $4.5$ 
simulated days.

As before, we observed how the values of \LearningRate{} and \DiscountFactor{} 
impact the quality of the measurements and the number of wireless 
transmissions in a sensor node.
However, differently from the synthetic scenarios, it is not possible to 
define the expected values in real world situations.
The main reason is that the environment is constantly changing and evolving,  
besides external factors that produce noise and change the environment 
itself.
Indeed, this is the core motivation of this work and what requires the
design of a solution that can adaptively adjust sensor nodes' sampling 
intervals.

\subsection{Number of transmissions}
\label{sec:num-tx}

\begin{figure}[t]
	\centering
	\includegraphics[width=0.45\textwidth]{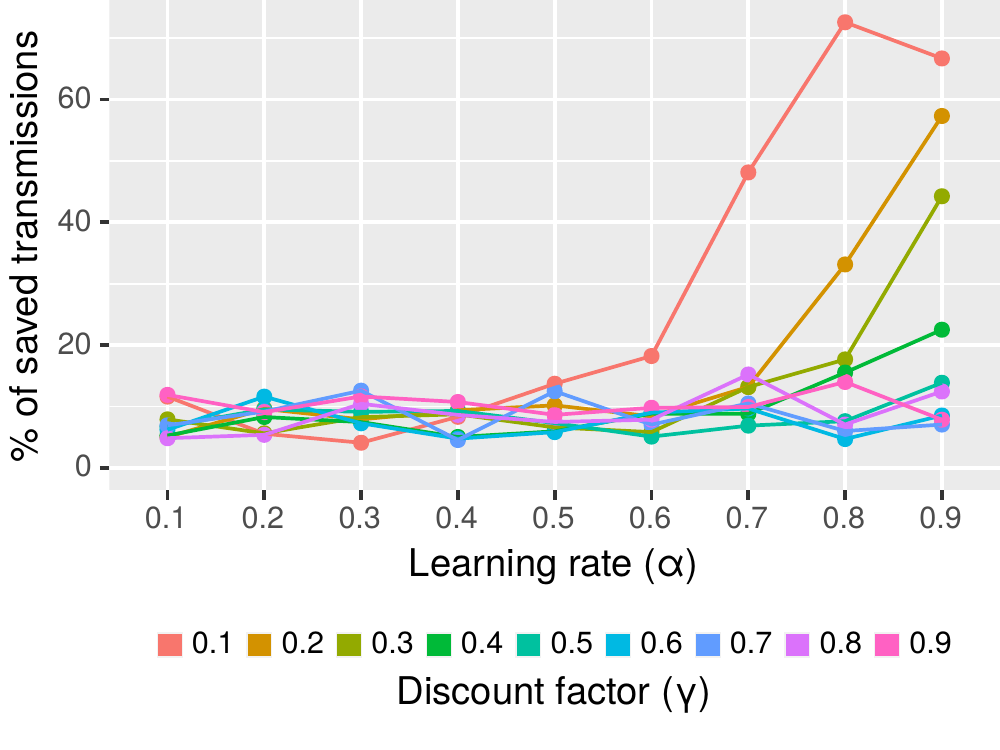}
	\caption{Impact of the Q-Learning agent in the reduction of the number 
of transmissions.}
	\label{fig:real-saved-tx}
\end{figure}

In our experiments, the number of transmissions in a sensor node achieves 
its maximum when the sampling interval is $30$ seconds and its minimum when 
the sampling interval is $240$ seconds.
Intuitively, setting the sampling interval to $60$, $120$ and $240$ seconds 
represents a reduction of respectively $50\%$, $75\%$ and $87.5\%$ in the 
maximum number of transmissions.

Figure~\ref{fig:real-saved-tx} illustrates how many transmissions could 
be saved when the Q-Learning was adopted to adjust the sensor node's sampling 
interval.
In this plot, we consider that the Q-Learning agent triggers one new 
transmission every time a new sampling interval is set.
In the best case, the number of transmissions can be reduced to up to $72.57\%$ 
of its maximum, when $\LearningRate{} = 0.8$ and $\DiscountFactor{} = 0.1$.
As observed in our preliminary results, the highest savings happen when 
\LearningRate{} is high (i.e., $\LearningRate{} \in \{0.7, 0.8, 0.9\}$) and 
\DiscountFactor{} is low (i.e., $\DiscountFactor{} \in 
\{0.1, 0.2, 0.3\}$).
That is, when the agent learns mostly from recent environment feedback and 
minimally from the expected \reward{}.
We highlight the importance of setting proper values to \LearningRate{} and 
\DiscountFactor{}, given that most of the cases do not reduce by more than 
$15\%$ the maximum number of transmissions.

\subsection{Efficiency}

\begin{figure}[t]
	\centering
	\includegraphics[width=0.45\textwidth]{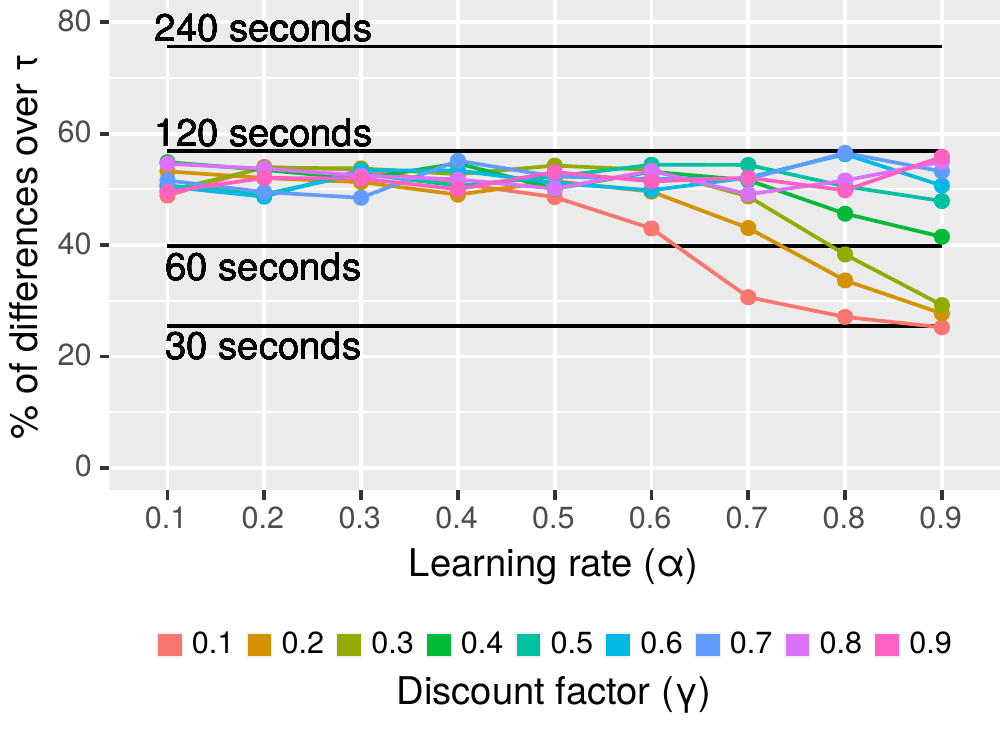}
	\caption{Percentage of consecutive measurements that differ by more 
than
\AcceptedThreshold{}. The black lines show the percentage observed the sampling 
intervals fixed to the values written in the plot.}
	\label{fig:real-statistics}
\end{figure}

Figure~\ref{fig:real-statistics} shows the percentage of consecutive 
measurements that differ by more than \AcceptedThreshold{}.
To help the understanding of the magnitude of the errors, we added four 
baselines that represent the rates that would be observed if the sampling 
intervals were fixed, based on the same data used in the simulations.
Again, the best results happen in scenarios using higher \LearningRate{} 
(i.e., $\LearningRate{} \in \{0.7, 0.8, 0.9\}$) and lower \DiscountFactor{} 
(i.e., $\DiscountFactor{} \in \{0.1, 0.2, 0.3\}$).

In the best result ($\LearningRate{} = 0.9$ and $\DiscountFactor{} = 0.1$), the 
rate of measurements over \AcceptedThreshold{} is similar to the 
scenario with a fixed sampling interval of $30$ seconds.
With the sampling interval set to $30$ seconds, we observed $16509$ pairs of
consecutive measurements that differed by more than \AcceptedThreshold{} with 
an average of $0.063^\text{o}$C.
Using Q-Learning, we observed $13679$ pairs of consecutive measurements that 
differed by more than \AcceptedThreshold{}, which differed by 
$0.078^\text{o}$C on average.

These small values strengthen the relevance of the reduction in the number of 
transmissions shown above, because they indicate that the avoided transmissions 
are, in fact, worthless in this scenario.
In conclusion, a real application that adopted Q-Learning with $\LearningRate{} 
= 0.9$ and $\DiscountFactor{} = 0.1$ would have saved around $65\%$ of its 
transmissions and observed an average of $0.024^\text{o}$C in the absolute 
difference between two consecutive measurements.

\section{Conclusion}
\label{sec:conclusion}

In this work, we adopted a Reinforcement Learning (RL) technique called 
\mbox{Q-Learning} to adjust the sampling interval of sensor nodes according 
to the expected changes in the environment. 
After explaining how the adaptive sampling can be formulated as a machine 
learning problem and showing that the environment's evolution can impact our 
algorithm, we evaluated the reduction in the number of 
transmissions and the quality of the reported data.
We presented the steps to setup the algorithm (i.e., \state{s}, \action{s}, and 
\reward{}) in a way that can be further exploited in several scenarios.

In our simulations, we could avoid nearly $73\%$ of transmissions 
in the best combination of parameters for the \mbox{Q-Learning} algorithm.
Observing the quality of the reported data, we noticed that the proposed
mechanism keeps similar quality to what would be observed if the smallest 
sampling interval was adopted.
Assuming that the radio transmissions are the main cause of energy 
consumption in monitoring WSNs, this solution may lead to significant savings 
in any scenario with wireless sensor nodes. 
Furthermore, the reduction in the number of transmissions can support WSNs to 
admit more sensor nodes, increasing their range and generating more knowledge 
about the monitored area.
To optimize resource usage, this mechanism can be further combined with other 
approaches, such as data aggregation and data compression.

We conclude that the use of an RL algorithm to control sensor
nodes' sampling intervals can be very profitable.
Furthermore,  we highlight that the proper choice of its parameters 
(\LearningRate{}~and~\DiscountFactor{}) can significantly impact the
results.
For instance, in our experiments, higher values of \LearningRate{} (i.e., 
$\LearningRate{} \in \{0.8, 0.9\}$) and lower values of \DiscountFactor{}  
(i.e., $\DiscountFactor{} \in \{0.1, 0.2\}$) provided the best cost-benefit 
in the ``saved transmissions''-``high quality'' relationship.
In our future work, we plan to implement the mechanism presented here 
to control the sampling interval of wireless sensor nodes in a real 
deployment.

\section*{Acknowledgment}

This work has been partially supported by 
the Catalan Government through the project SGR-2014-1173 and by the 
European Union through the project FP7-SME-2013-605073-ENTOMATIC.

\bibliographystyle{IEEEtran}
\bibliography{IEEEabrv,bibliography}
 
\end{document}